\documentstyle[12pt, preprint] {aastex}
\newcommand{\hm}{H$_{2}$}

\begin{document}

\title{Interstellar ices as witnesses of star formation: selective deuteration of water and organic molecules unveiled.}
\author{S. Cazaux\altaffilmark{1}}
\affil{Kapteyn Astronomical Institute, PO box 800, 9700AV Groningen, The Netherlands}

\author{P. Caselli\altaffilmark{2}}
\affil{School of Physics and Astronomy, University of Leeds, LS2 9JT, Leeds, UK}

\author{M. Spaans\altaffilmark{1}}
\affil{Kapteyn Astronomical Institute, PO box 800, 9700AV Groningen, The Netherlands}

\authoremail{cazaux@astro.rug.nl}

\begin{abstract}
{Observations of star forming environments revealed that the abundances of some deuterated interstellar molecules are markedly larger than the cosmic D/H ratio of 10$^{-5}$. Possible reasons for this pointed to grain surface chemistry. However, organic molecules and water, which are both ice constituents, do not enjoy the same deuteration. For example, deuterated formaldehyde is  very abundant in comets and star forming regions, while deuterated water rarely is. In this article, we explain this selective deuteration by following the formation of ices (using the rate equation method) in translucent clouds, as well as their evolution as the cloud collapses to form a star. Ices start with the deposition of gas phase CO and O onto dust grains. While reaction of oxygen with atoms (H or D) or molecules (H$_2$) yields H$_2$O (HDO), CO only reacts with atoms (H and D) to form H$_2$CO (HDCO, D$_2$CO). As a result, the deuteration of formaldehyde is sensitive to the gas D/H ratio  as the cloud undergoes gravitational collapse, while the deuteration of water strongly depends on the dust temperature at the time of ice formation. These results reproduce well the deuterium fractionation of formaldehyde observed in comets and star forming regions and can explain the wide spread of deuterium fractionation of water observed in these environments.}\rm

\end{abstract}


\section{Introduction}
Stars form through the collapse of interstellar clouds which are composed of gas and dust. As the cloud collapses, its density increases, its temperature decreases and it becomes shielded from external UV radiation that would otherwise dissociate chemical species present in the medium. In these regions, dust grains grow thick icy mantles from the deposition \rm{of atomic and}\rm\ molecular species onto their surfaces. This has been confirmed by observations of starless cores (\citealt{tafalla2006}), which show that most gas phase species suffer a significant drop towards the core center. The missing gas species constitute the icy mantles that cover dust grains.  
As a star forms and heats up its environment, \rm{ a portion of the}\rm\ icy mantles are released into the gas phase. This phase of star formation, called the hot core (massive stars) or hot corino (low mass stars) phase, exhibits a very complex chemistry rich in oxygen and nitrogen bearing molecules (\citealt{cazaux2003}), but also shows species such as H$_2$CO and methanol, which are building blocks for more complex organic molecules essential for the formation of life. Also, while deuterium is known to be 10$^{5}$ times less abundant than hydrogen (\citealt{linsky2003}), many of these environments present  a chemistry rich in highly deuterated species. 
\rm{In particular, deuterated methanol has been found in molecular clouds (\citealt{parise2004}) and in low mass protostars, with a maximum of  $\sim$ 30$\%$ for CH$_2$DOH/CH$_3$OH  (\citealt{parise2002}) and of $\sim$ 3$\%$ for CD$_3$OH/ CH$_3$OH (\citealt{parise2004}), while no detection has been reported in comets (\citealt{crovisier2004}). Formaldehyde present a high deuterium fractionation as well (see Table 1), HDCO/H$_2$CO is of a few percent towards protostellar cores (\citealt{roberts2007}, \citealt{bacmann2002}) and towards hot cores (\citealt{turner1990}) and of a tenth of a percent towards hot corinos (\citealt{parise2006}).}\rm\ Its doubly deuterated form D$_2$CO/H$_2$CO can reach up to 5$\%$ in hot corinos (\citealt{parise2006}). Species that have been observed with 3 deuterium atoms (ND$_3$; \citealt{vanderTak2002}) and CD$_3$OH (\citealt{parise2004}) are expected to be 10$^{15}$ times less abundant than their hydrogen form, \rm{with respect to elemental abundances.}\rm\ Yet, their enhancement can be of 12 orders of magnitude (\citealt{ceccarelli2007}).

This high degree of deuteration is attributed to grain surface chemistry. However,  while water co-exists with methanol and formaldehyde in ices, it rarely \rm{has}\rm\ the same deuteration. In general, the degree of deuteration of organic material is very high compared to that measured in water (\citealt{ehrenfreund2002}). The deuterium fractionation of water (Table 1) can be about  HDO/H$_2$O$\sim$0.01$\%$ towards massive hot cores (\citealt{gensheimer1996}) and $\sim$0.02-0.05$\%$ in comets  and asteroids (\citealt{altwegg2003}). These values are similar to SMOW (Standard Mean Ocean Water), which reports a fraction HDO/H$_2$O  of 0.015$\%$ in our oceans. \rm{However,  some detections showed a high HDO/H$_2$O ratio of few $\%$ towards the hot corino NGC1333-IRAS2A (\citealt{liu2011}), and IRAS16283-2422 (Coutens et al. in prep., \citealt{parise2005}). }\rm 
In this study, we address the selective deuteration of water compared to formaldehyde. We used H$_2$CO as a representative of organic molecules (H$_2$CO is a building block of most of the complex organic molecules found in space), but our study also apply to other species formed with CO, such as CH$_3$OH. First, we determine the chemical gas phase content of the region where ices form.  Then, we use a rate equation method that follows the accretion and reactions of gas phase species that yield ices. Finally, we simulate how the ice compositions change as they undergo gravitational collapse to form stars. These latest results are compared with observations of star forming environments as well as cometary material.

\section{Origin of the ices}
Ices are formed in translucent clouds \rm(clouds where \hm\ dominates over HI and carbon is converted from C II to C I to CO)\rm, in regions that are shielded from UV radiation. At extinctions Av$\ge$3 mag, species from the gas phase accrete onto dust grains and initiate the formation of ices (\citealt{whittet2001}, \citealt{chiar2011}). Therefore, the chemical content of the gas phase at this extinction is crucial to determine which species accrete onto dust and what the composition of ices is. Gas phase chemistry of dense cores (\citealt{roberts2003}) shows that a high D/H ratio can be reached because of the inclusion of ions such as HD$_2^+$ and D$_3^+$ in gas phase chemistry, which produce deuterons upon dissociative recombination. This gas phase model is able to reproduce a high deuteration of formaldehyde in dense cores. Also, the degree of deuteration of molecular ions are sensitive to the ortho:para H$_2$ ratio and hence to the chemical and thermal history of the precursor molecular cloud (\citealt{flower2006}). 

In this study, we concentrate on the ice formation driven by the accretion of gas phase species. \rm{To this effect, we determine the gas phase composition (in hydrogen and deuterium) of a medium submitted to UV radiation from neighboring stars, by using a model that couples gas phase and grain surface chemistry  (\citealt{cazaux2009}). This model follows the gas phase densities of molecular and atomic hydrogen as well as deuterium, as the medium becomes shielded to the external UV field (\citealt{draine1996}). The results reported in figure 1 show the chemical composition in hydrogen and deuterium as function of the extinction, and therefore provide the initial conditions that are used in the next section (table 2) for the formation of ices (at A$_V$$\sim$3). Note that neither oxygen, CO and ice chemistry are treated here, our goal being to assess the hydrogen and deuterium content of a medium in the region where ices are forming.}\rm\ As the hydrogen becomes molecular, deuterium can remain atomic, which leads to an enhanced D/H ratio at low extinctions. Then, when the medium is shielded, the chemistry lead by ions drives a D/H ratio that can reach 0.3 (\citealt{roberts2003}). In Figure 1, the H/H$_2$ and D/HD fronts are presented, as well as the D/H ratio as function of the extinction for an UV radiation field  similar to the one in our local interstellar medium (defined as G$_0$=1; \citealt{habing1968}). The D/H ratio increases with density for extinctions higher than 1 mag. 
In this study we consider the formation of ices in translucent clouds with initial gas phase conditions listed in Table 2.

\section{Dust favors deuteration}
In our grain surface chemistry model, we consider the formation of ices by accreting species from the gas phase, and let the icy mantles grow until the CO and oxygen from the gas phase are strongly depleted onto dust. Jeans instability, turbulence and/or triggering due to winds from nearby young stars will then lead to local density increase and gravitational collapse of a cloud fragment.  During the collapse, the molecules remaining in the gas can still interfere with the ices and change their composition. Our model is therefore separated into two parts: 1) the slow formation of ices in diffuse environments;  2) the gravitational collapse of dense icy regions of the cloud beginning star formation.

\subsection{Translucent clouds}
\subsubsection{Gas phase composition}
Ices form in shielded regions (A$_V$$\sim$3 mag), with CO and oxygen equally abundant in the gas phase (\citealt{hollenbach2009}) and scale as 1.5$\times$ 10$^{-4}$ n$_H$, where n$_H$ the total hydrogen density. We assume that ices are first formed in translucent clouds, with initial gas phase conditions listed in Table 2. As time passes, oxygen and CO will mostly be depleted onto dust, impoverishing the gas phase. Their densities can be calculated as:
\begin{equation}
\frac{d n_O }{dt}=1.5\times 10^{-4} n_H -O_{dust} \times 4 n_{dust} \sigma n_{site},
\end{equation}
\begin{equation}
\frac{d n_{CO}}{dt}=1.5\times 10^{-4} n_H -CO_{dust} \times 4 n_{dust} \sigma n_{site},
\end{equation}
where O$_{dust}$ and CO$_{dust}$ are the amount of oxygen and CO on the dust, in monolayers, $n_{sites}\sim$10$^{15}$ cm$^{-2}$ is the density of sites, n$_{dust}$ the density of dust (cm$^{-3}$) and $\sigma$ their cross section (cm$^{-2}$).  Here we consider a grain size distribution that takes into account very small grain particles (\citealt{weingartner2001}), and derive n$_{dust} \sigma$/n$_H$=2.8 10$^{-21}$ cm$^{2}$. With this, we compute the densities of gas phase CO and oxygen with time, and also follow in parallel the chemistry occurring on the dust surfaces.

\subsubsection{Grain surface chemistry}
The gas phase composition of the environments where ices originate, discussed above, is reported in table \ref{para}.  We follow the population of species on the dust by using the rate equation method. Species present in the gas accrete onto dust at a rate:
\begin{equation}
R_{acc_i} = \frac{n_i v_i  S}{n_{site}} \rm{monolayer\ s^{-1}},
\end{equation}
where n$_i$ and v$_i$ are, respectively, the densities and velocities of the species i, and $S$ is the sticking coefficient of the species with the dust (here we consider S=1).  Since our rate equation model follows the populations of the species of the surface (with 1 monolayer =100$\%$ coverage), this rate is expressed in monolayer s$^{-1}$. The binding energies between the chemical species and the grain surface are assumed to be weak (Van der Waals  or physisorption), which is typical for icy surfaces. These energies are taken from previous studies (\citealt{cuppen2007}), with the exception of oxygen, that we take as 1390 K (\citealt{bergeron2008}) and H$_2$ as 520K (\citealt{dulieu2005}). The binding energies of CO and hydrogenated/deuterated forms are all considered to be 1200 K  (\citealt{allouche1998}).

The species $i$ present on the surface may return to the gas phase if they evaporate with a rate:
 \begin{equation}
R_{evap(i)} = \nu_i \times \exp{(-\frac{E_i}{k_BT})} ,
\end{equation}
where $\nu_i$ is the oscillation factor  of the species i (which is of 10$^{12}$ s$^{-1}$ in physisorbed sites) and E$_i$ is the binding energy of the species $i$, as reported in \cite{cazaux2010} and discussed above.
 
On the surface, atoms and molecules can travel on the dust with a mobility (\citealt{cazaux2004}, \citealt{cazaux2010}):
\begin{equation}
R_{i} = \nu_i \times exp({-\frac{E_{pp}}{k_B T}}) 
\end{equation}
where E$_{pp}$ is the energy of the barrier between two physisorbed sites (considered as 2/3 of the binding energies E$_i$).

Species on the surface can be photodissociated by UV photons (\citealt{cuppen2007}), and can meet another species to form a product that will either stay on the surface or be released on the gas phase (\citealt{cazaux2010}). In this study, we follow the coverage of the different species in order to determine the ices composition.

We have constructed a chemical network that considers the hydrogenation and deuteration, but also the H-D and D-H exchanges for H$_2$CO, HDCO and D$_2$CO. The most important reactions of the chemical network involving O and CO are presented in Figure 2 with the associated barriers for the reactions to occur, E$_b$. \rm{The characteristic time for a reaction can be calculated as $\sim$ $10^{12} \times \exp(-0.406\times \sqrt{\mu_{red}\times E_b})$ (oscillation factor times  the probability of tunneling through a barrier of 1\AA) , with $\mu_{red}$ the reduced mass for the reaction. This reaction occurs if its characteristic time is smaller than the time for the species to leave the site. }\rm

The chemistry that involves oxygen has been described in a previous work (\citealt{cazaux2010}). For the chemistry of CO, the different barriers for the reactions considered in our model have been derived by several authors (\citealt{fuchs2009}, \citealt{hidaka2007}, \citealt{hidaka2009}). The reactions H$_2$ + CO $\rightarrow$ H$_2$CO as well as HCO + H$_2$ $\rightarrow$ H$_2$CO + H, have very high barriers of  38400 K and 9000 K (\citealt{manion2008}), respectively.  Also, the reactions involving D$_2$ are negligible (\citealt{hidaka2009}). Therefore, we do not include reactions involving H$_2$, HD and D$_2$  in the CO chemistry network. 

Recent  studies (\citealt{ratajczak2009}, \citeyear{ratajczak2011}) examined exchanges between H and D in ices composed of water and deuterated methanol as these ices warm up. D/H exchanges  concerns the hydroxyl functional group of methanol and therefore occur between CD$_3$OD and H$_2$O that lead to CD$_3$OH and HDO. In the ices, the amount of CD$_3$OD is very low compared to the amount of water. These H/D exchanges  can be responsible  for a very small enhancement of HDO and are therefore not considered in this study.



\subsection{Collapse and star formation}

In the second phase of our model, the dust grains covered by ices are present in a denser environments that undergoes gravitational collapse. We assume the density of the medium to be n$_H$$\sim$10$^5$ cm$^{-3}$ and our model cloud to collapse at the free-fall rate. The simulations follow the gas density evolution as:
\begin{equation}
\frac{d n_H }{dt}=\frac{ n_H}{t_{ff}},
\end{equation}
where $t_{ff}$=$\sqrt{3\pi/32G\rho} \sim 10^5$ yr is the free-fall time, with G the gravitational constant and $\rho$ the mass density. Because CO and oxygen are mostly depleted onto the dust, only H, D, H$_2$ and HD can accrete onto the icy surfaces with an atomic D/H ratio $\sim$ 5 10$^{-2}$. The initial gas phase composition of this step of our simulation is reported in Table 2. \rm{As the cloud collapses, and the medium becomes denser, the D/H ratio increases as shown in Figure 1. }\rm Molecular hydrogen, which is by far the most abundant molecule in the medium,  accretes onto the last layer of ice. Once one layer of H$_2$ covers the ice (around n$_H$$\sim$10$^6$cm$^{-3}$), we assume that no further accretion is possible since the binding energies of species with H$_2$ are very low, and therefore species bounce back into the gas phase. The ice constituents are followed during the collapse of the cloud, as a function of the density of the medium.

\section{Results}
We have computed the abundance of water and formaldehyde as well as their deuterated forms in interstellar ices.

The oxygen accreting onto the dust transforms into water ice. Fig~3a shows the formation of H$_2$O and HDO on the dust surface as a function of time. The deuteration of water is very sensitive to the dust temperature. At low T$_{dust}$ (12~K),  H$_2$  and H are both present on the ices, and oxygen can react both with H and H$_2$. Because H$_2$ is much more abundant than H, water forms through the reaction O+H$_2$ while the formation of HDO involves atomic D. Therefore, the ratio HDO/H$_2$O (fig~3b) scales with D/H$_2$. \rm{At T$_{dust}$= 15~K, short time scales allow a important amount of H atoms to be ice constituents, and oxygen reacts with atomic H to form water. As time proceeds, H atoms disappear by reacting with species present in the ices and H$_2$ becomes largely dominant. Oxygen atoms coming from the gas phase then react predominantly with H$_2$ molecules. }\rm Therefore, the ratio HDO/H$_2$O scales with D/H for short time scales and with D/H$_2$ for long time scales. At T$_{dust}$=17~K, H$_2$ molecules evaporate too fast to allow the reaction O+H$_2$ to occur. Water is formed through successive hydrogenation of oxygen, and the ratio HDO/H$_2$O scales with D/H. 

The CO accreting onto the dust transforms into formaldehyde. Fig.~3c shows the build up of H$_2$CO, HDCO and D$_2$CO ices. The formation of formaldehyde is made through the reactions CO+H$\rightarrow$HCO and HCO+H$\rightarrow$H$_2$CO. The CO-ices build up slowly, and CO present on the dust associates with an incoming H atom, or with a D atom. Therefore, HDCO/H$_2$CO and D$_2$CO/H$_2$CO (fig~3d) scale with D/H and (D/H)$^2$, respectively.

Once the medium becomes denser, the cloud made of gas and dust (covered by ices) can undergo gravitational collapse and form a star. This star heats up its environment and the content of the ices is released into the gas phase (hot core/corino). At later stages, the star is surrounded by a disk, and the content of the ices can be present in the warmer parts, or can be locked into cometary material. In these different phases of star formation, deuterated water and formaldehyde have been observed.

\rm{Deuterium fractionation of formaldehyde and water has been observed toward several astrophysical objects, as reported in table 1. While HDCO/H$_2$CO and D$_2$CO/H$_2$CO ratios range between 0.01-0.28 and 0.01-0.04, respectively, HDO/H$_2$O ratio are much more spread, ranging from few 10$^{-4}$ to 0.02. In Fig~3e and Fig~3f, we compare the results of our model to these observations. }\rm

Fig.~3f, presents the evolution of the deuterated forms of formaldehyde as the cloud collapses. As the ices are present in a denser environment with a higher D/H ratio, the fraction of deuterated formaldehyde increases by 1 order of magnitude. Our model reproduces the high degree of deuteration of formaldehyde as the cloud collapses. Also, the fractionation of formaldehyde is not sensitive to the temperature at which the ices formed in the translucent clouds. 

For water, on the other hand, Fig.~3e shows that the ratio HDO/H$_2$O is constant as the cloud collapses. Once H$_2$O and HDO are formed, a very important barrier (9600K) has to be overcome to break these molecules. Therefore, the deuterium fractionation is set during the formation of ices, in the translucent clouds, and deuterated water retains the memory of the ice formation. The ratio HDO/H$_2$O is extremely sensitive to the dust temperature at which the ices formed in the translucent clouds. At low temperatures, water is formed with O+ H$_2$, while at higher temperatures H$_2$ evaporate, and water is mode through O+H. As a result, \rm{our model can explain the spread in the HDO/H$_2$O ratio observed in different astrophysical objects.}\rm 

\rm{The mantles species that are released in the gas phase could be transformed in other species through gas phase reactions. \cite{goumans2011} showed that gas phase reactions of H$_2$CO with H(D) would lead to HCO + H$_2$(HD), while H +D$_2$CO would lead to CHD$_2$O. In this case, the deuterium fractionation of formaldehyde could be increased by further reactions in the gas phase. However, \cite{charnley1997} show that the HDO/H$_2$O ratio stays constant after ices evaporate, while HDCO/H$_2$CO could be increased by a factor 5. Therefore, the deuterium fractionation found in this study, that is the result of the chemistry occurring in the ices, should not be dramatically changed by the gas phase chemistry that follows ice evaporation}.\rm

\section{Conclusions}
We used the rate equation method to simulate the formation of water and formaldehyde, as well as their isotopologues on dust grain surfaces. We model the formation of ices in translucent clouds and incorporate these ices in denser environments that undergo gravitational collapse to form stars. 

We find that  HDCO/H$_2$CO and D$_2$CO/H$_2$CO depend on the gas D/H ratio which increases as the medium collapses and becomes denser.

For water, on the other hand, HDO/H$_2$O strongly depends on the dust temperatures at which ices form since water formation involves H$_2$ at low T$_{dust}$ (H$_2$ very abundant on dust surface) and H at higher T$_{dust}$ (H$_2$ evaporate). Consequently, the degree of water deuteration is directly related to the physical conditions (i.e., dust temperature) at which ices form in the quiescent cloud just before it collapses to form stars. \rm{H$_2$CO is the simplest representative of organic species formed via hydrogenation of CO onto dust surfaces. Our results can then be generalized to other species (such as CH$_3$CO), which also depends on surface hydrogenation of CO. This explains the selective deuteration of water and organic molecules observed in the ISM, similar to that found in comets, IDPs and chondrites (\citealt{ehrenfreund2002}, Fig. 9)\rm.

\subsection*{ACKNOWLEDGMENTS}
The authors wish to thank the anonymous referee for her/his constructive comments. S. C. is supported by the Netherlands Organization for Scientific Research (NWO). 

\clearpage

\begin{table}
\caption{Deuterium fractionation of formaldehyde and water observed in different environments \label{obs}}
\begin{tabular}{l|l|l|l|l}
Environment & hot corinos & hot cores & compact ridge &  comets\\ \hline
Densities (cm$^{-3}$)& $\sim$10$^8$$^a$ &10$^9$- 10$^{10}$$^b$ &10$^6$$^c$ & 10$^{10}$$^d$ \\ \hline
HDCO/H$_2$CO& 0.07-0.22$^e$&0.01-0.03$^e$&0.09-0.26$^f$&$\le$0.05$^g$ 0.28$^h$\\
D$_2$CO/H$_2$CO& 0.01-0.03$^e$&$\le$0.01$^e$&0.016-0.03$^f$&\\
HDO/H$_2$O&$\le$6 10$^{-4}$$^i$&6 10$^{-5}$-5 10$^{-4}$$^l$&&3 10$^{-4}$-4 10$^{-4}$$^m$\\ 
&$\ge$0.01$^j$&&&\\ 
&0.007-0.027$^k$&&&\\ 
\multicolumn{5}{l}{$^a$ \cite{cecarelli2005}, $^b$\cite{roberts2007}, $^c$  \cite{charnley1997},$^d$\cite{woitke2009}}\\
\multicolumn{5}{l}{$^e$ \cite{roberts2007}, \cite{parise2006}, $^f$\cite{turner1990}}\\
\multicolumn{5}{l}{$^g$  \cite{crovisier2004} Hale-Bopp,$^h$\cite{kuan2008} C/2002 T7}\\
\multicolumn{5}{l}{$^i$\cite{liu2011} NGC1333-IRAS2A, $^j$\cite{jorgensen2010} NGC 1333-IRAS4B}\\
\multicolumn{5}{l}{$^k$Hershel results IRAS16293-2422 Coutens et al. in prep.}\\
\multicolumn{5}{l}{$^l$\cite{gensheimer1996} , $^m$\cite{meier1998},\cite{villanueva2009}}
\end{tabular}
\end{table}

\begin{figure}
\plottwo{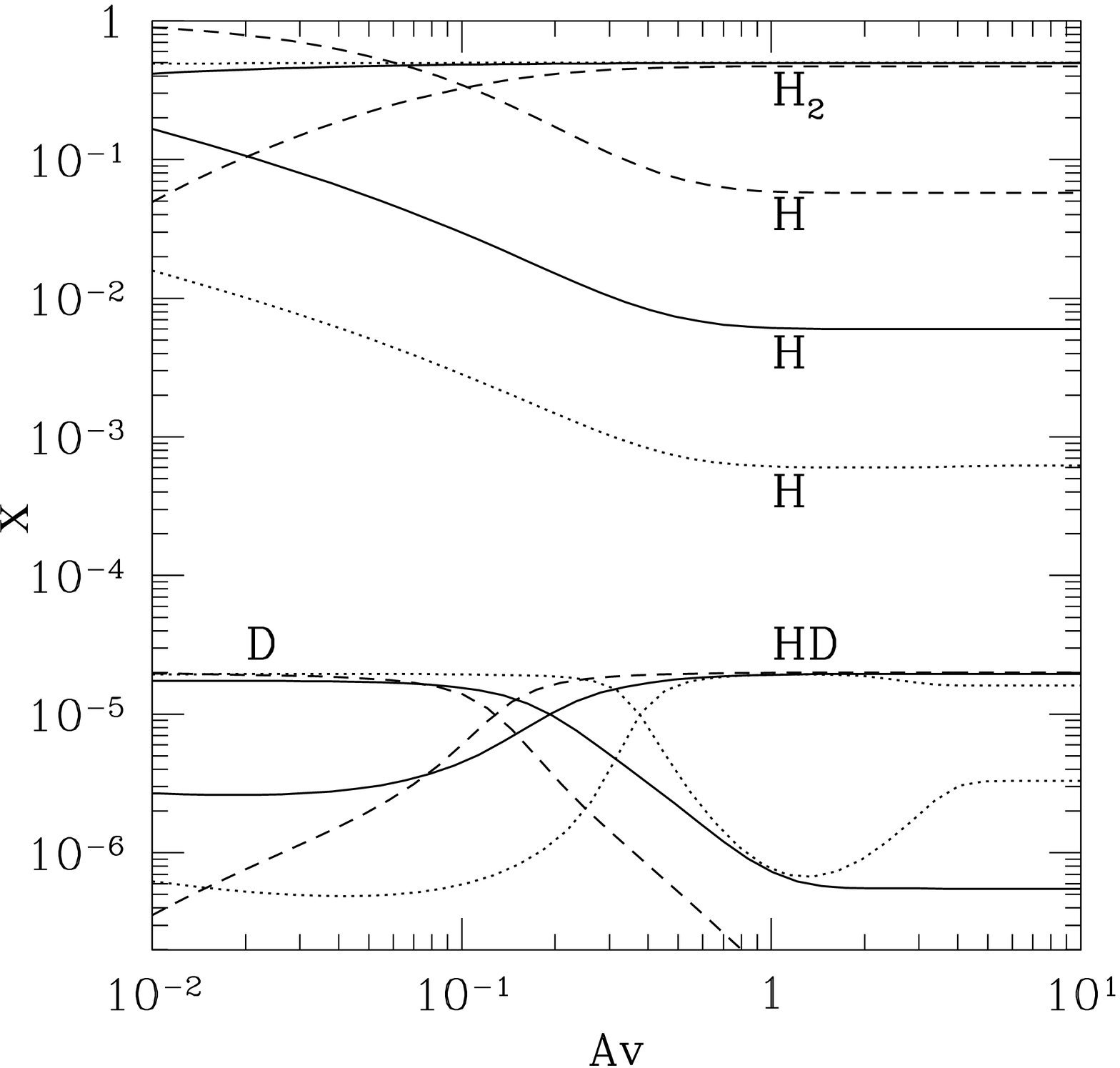}{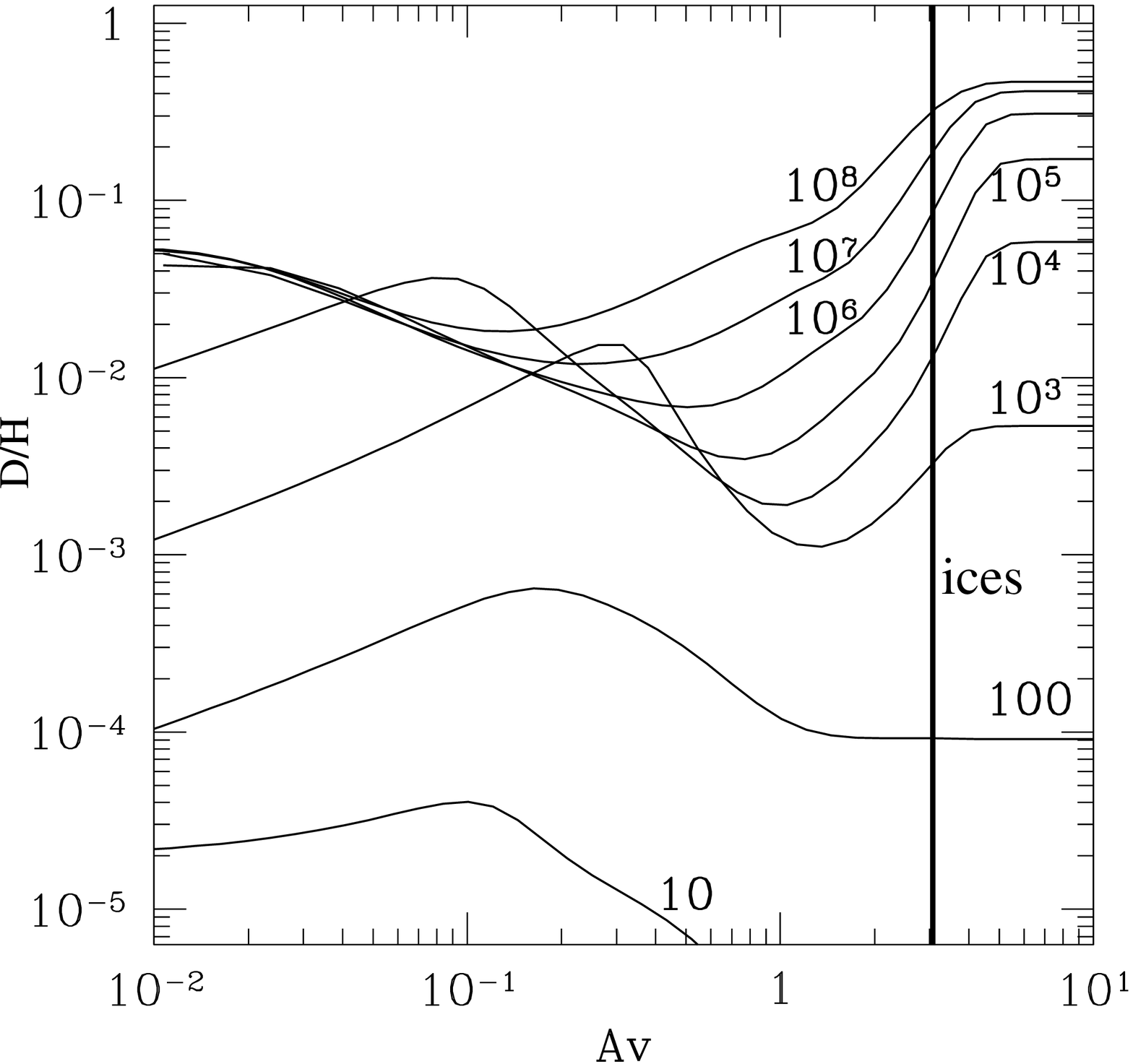} 
\caption{Left panel: atomic (H and D) and molecular (H$_2$ and HD) hydrogen and deuterium abundances with respect to total hydrogen (nHI+nH$_2$)  in a cloud with a density of 10 (dashed), 100 (solid) and 10$^3$ cm$^{-3}$ (dotted) as a function of the visual extinction (mag). Right panel: Variation of the D/H ratio as a function of the visual extinction (mag) for densities varying from 10 to 10$^8$ cm$^{-3}$. The line shows A$_V$$\sim$3, where the ices initiate.\label{DH}}
\end{figure}

\begin{table}
\caption{Initial conditions of translucent (at A$_V$$\sim$3) and collapsing clouds used in our simulations \label{para}}
\begin{tabular}{l|l|l|l|l|l|l|l|l}
Env. &n$_{\rm{H}}$&n$_{\rm{HI}}$ &n$_{\rm{H_2}}$ & n$_{\rm{DI}}$&n$_{\rm{OI}}^a$ &n$_{\rm{CO}}^a$ &T$_{\rm{dust}}^b$&T$_{\rm{gas}}^b$\\ \hline
Translucent& 10$^3$ & 0.5 & 5 10$^2$ & 3 10$^{-3}$  & 0.15 & 0.15  &12& 20\\
& 10$^3$ & 0.5 & 5 10$^2$ & 3 10$^{-3}$ & 0.15 & 0.15  &15& 30\\
& 10$^3$& 0.5 & 5 10$^2$ & 3 10$^{-3}$ & 0.15  & 0.15 &17& 70\\
Collapsing& 10$^5$ & 0.5& 5 10$^4$ & 2.5 10$^{-2}$ & 0.001 & 0.001 &12& 12\\
\multicolumn{9}{l}{n densities in cm$^{-3}$, n$_{\rm{H}}$ total hydrogen density (n$_{\rm{HI}}$ + n$_{\rm{H_2}}$) }\\
\multicolumn{9}{l}{T temperatures in K; $^a$ \citealt{hollenbach2009}, $^b$ \citealt{cuppen2007}}
\end{tabular}
\end{table}

\begin{figure}
\plottwo{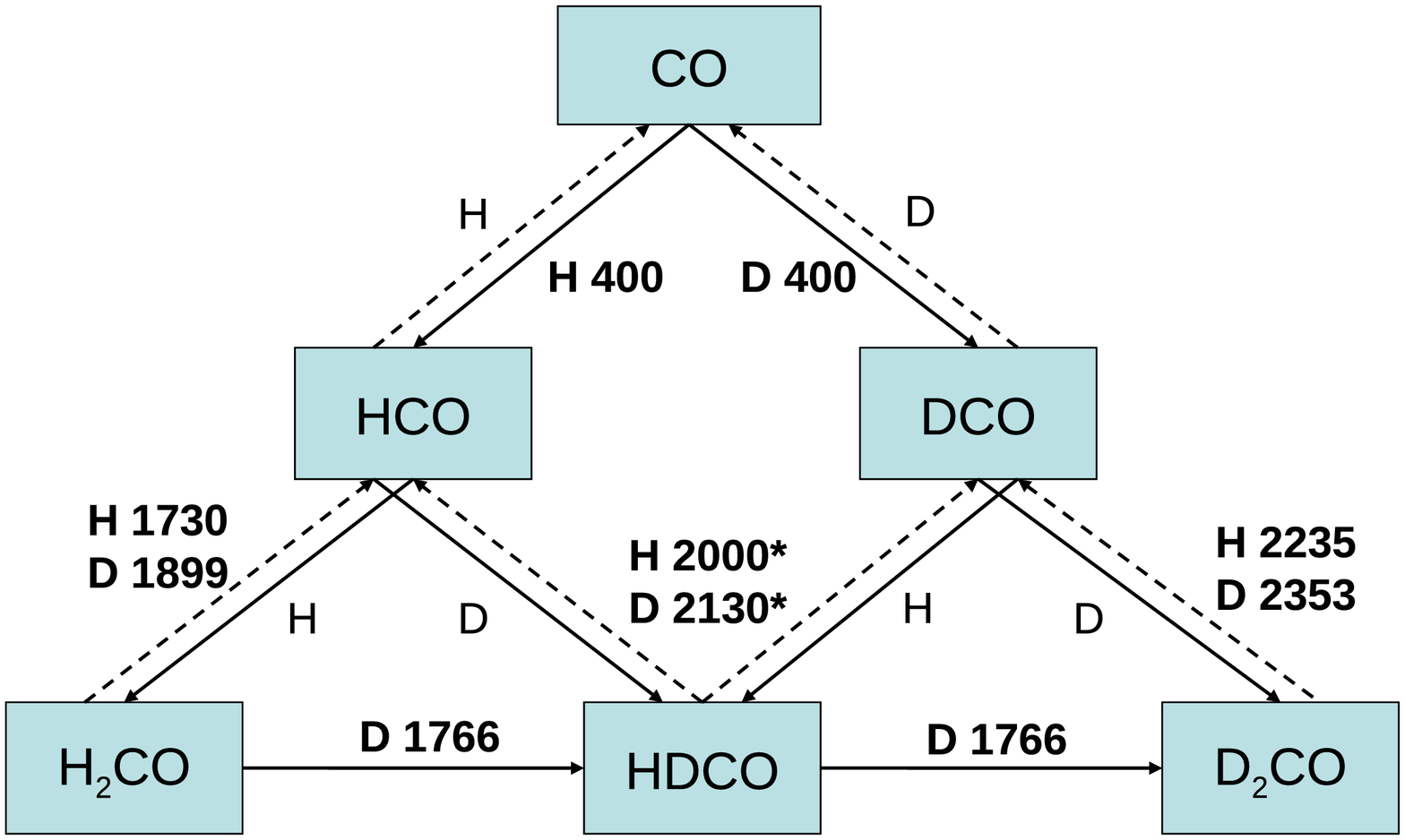}{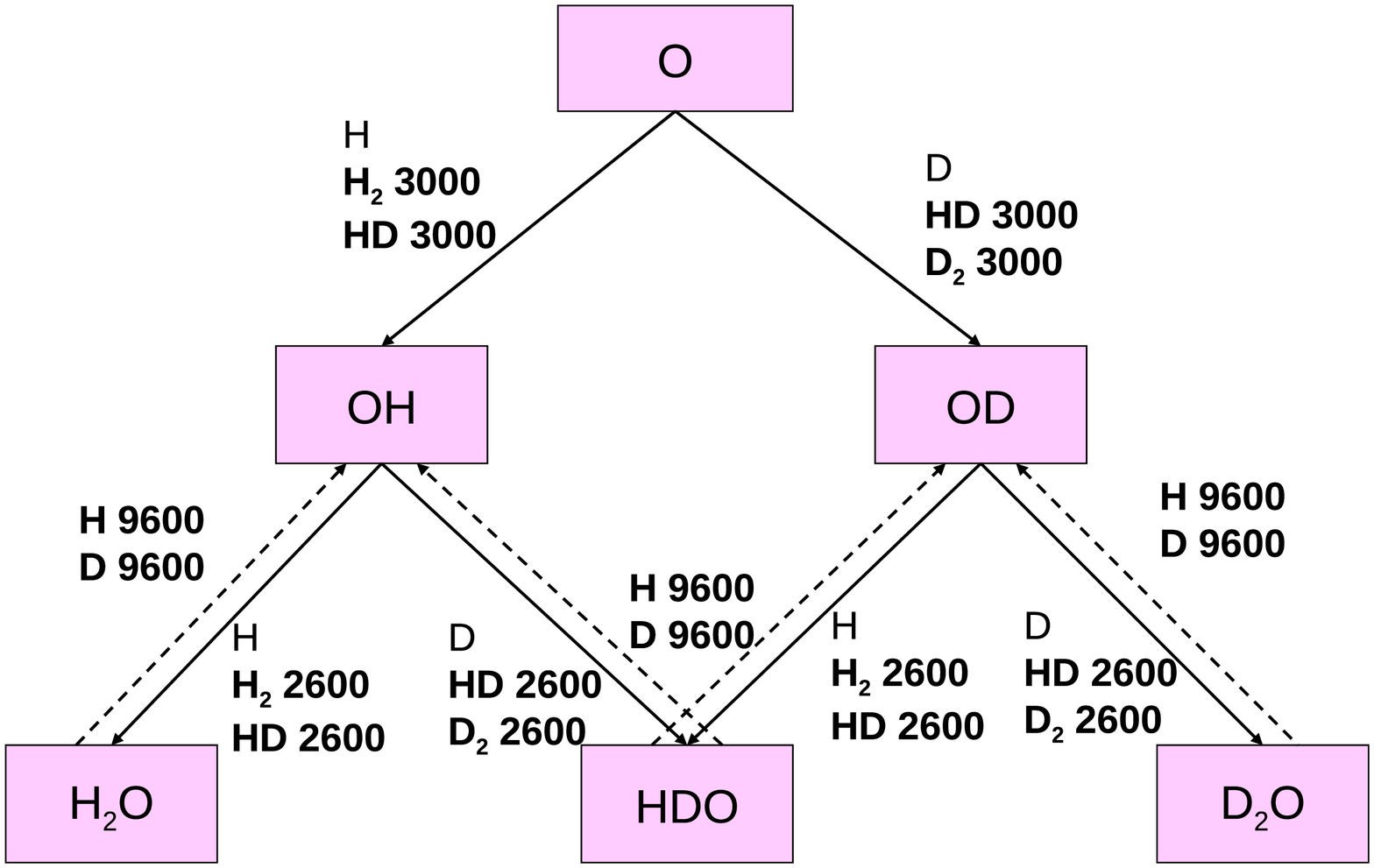} 
\caption{Principal reactions involving CO (left) and O (right) that are considered in our network. The reactions in boldface are associated with barriers (in Kelvin).  Barriers that have not been determined, but that we consider as the mean of similar reaction are denoted with an asterisk (*). Note that O reacts with atoms and molecules, while CO only reacts with atoms.}
\label{oco}
\end{figure}

\begin{figure}
\epsscale{0.87}
\plottwo{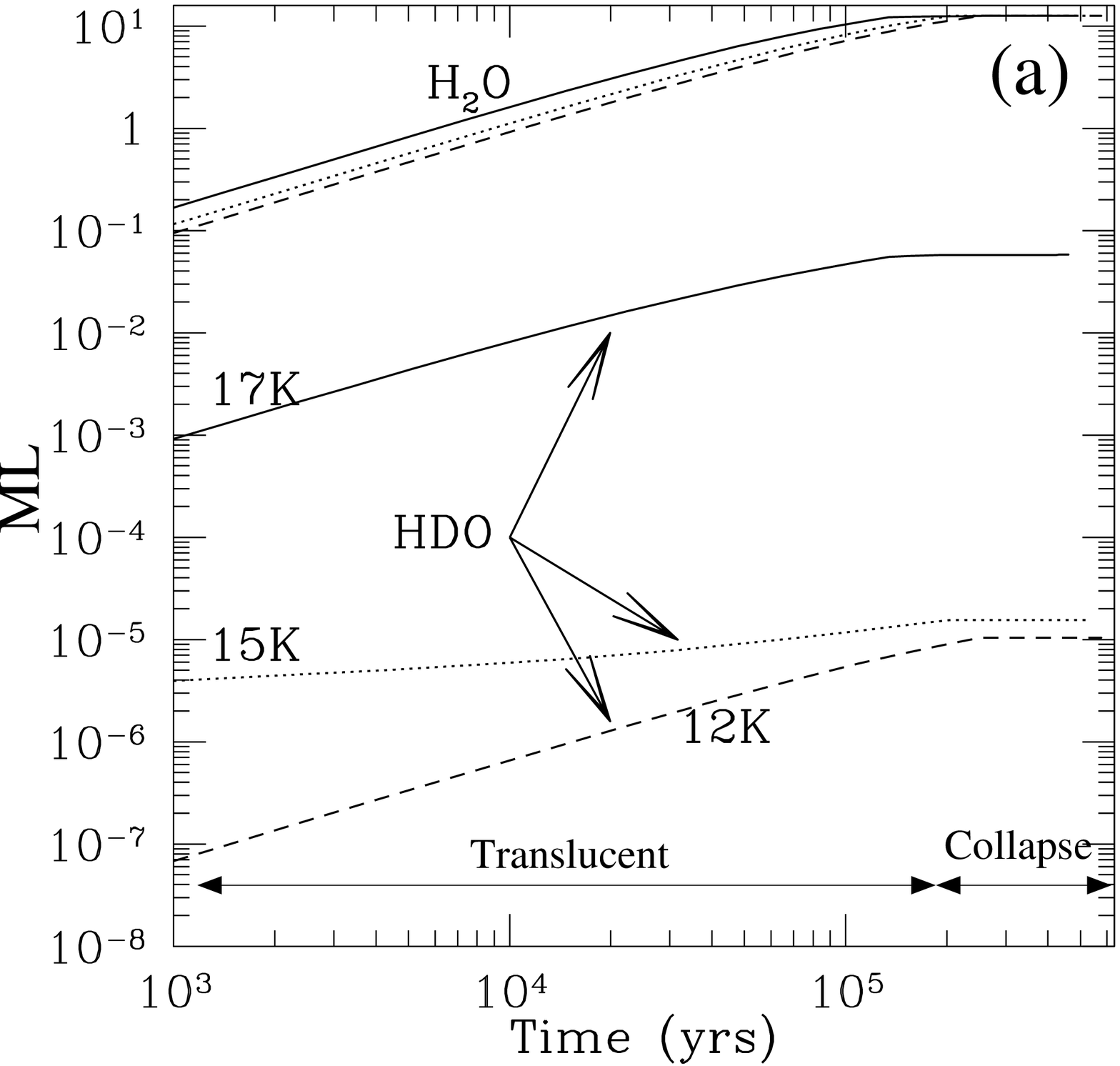}{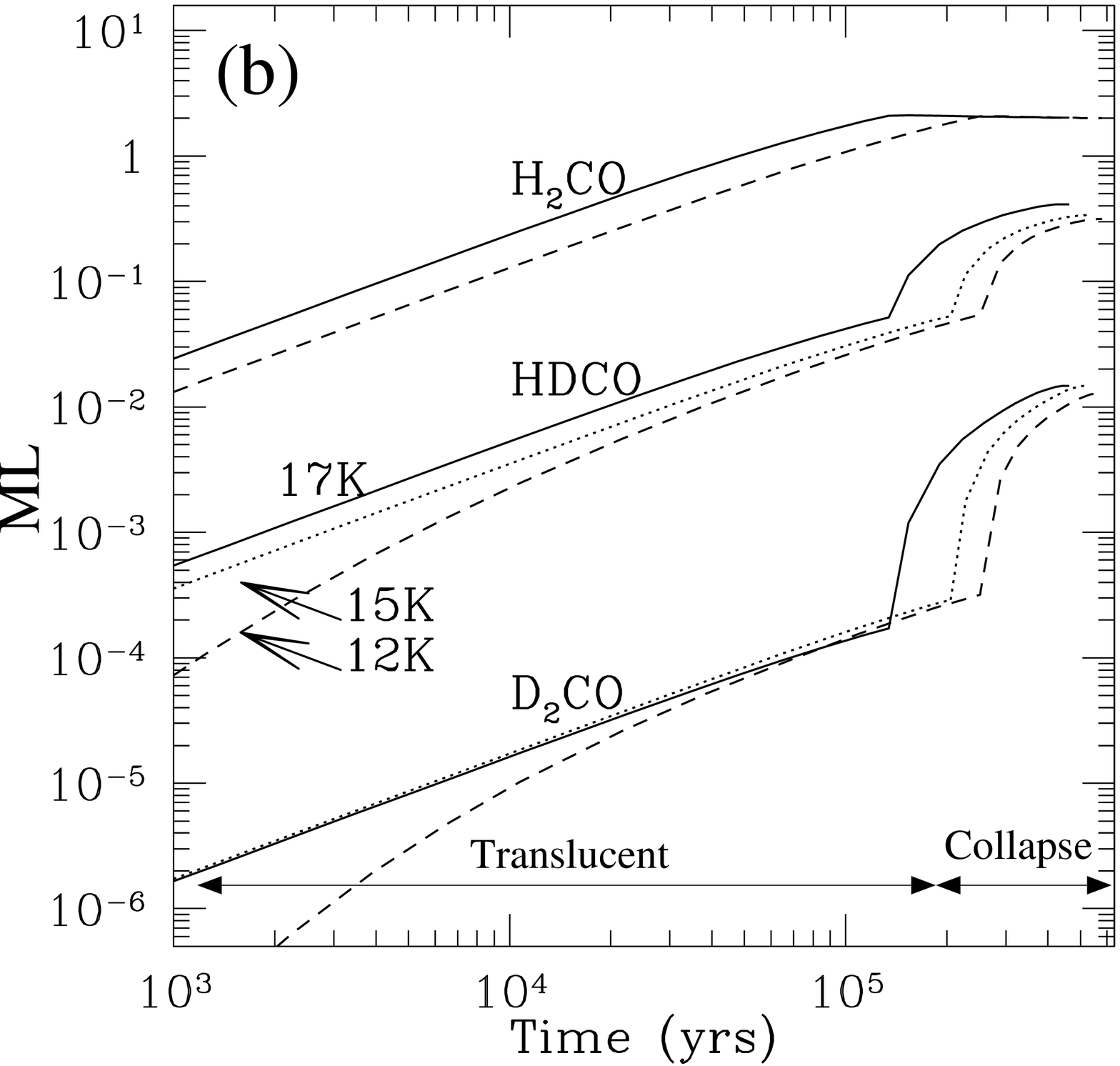} \\
\plottwo{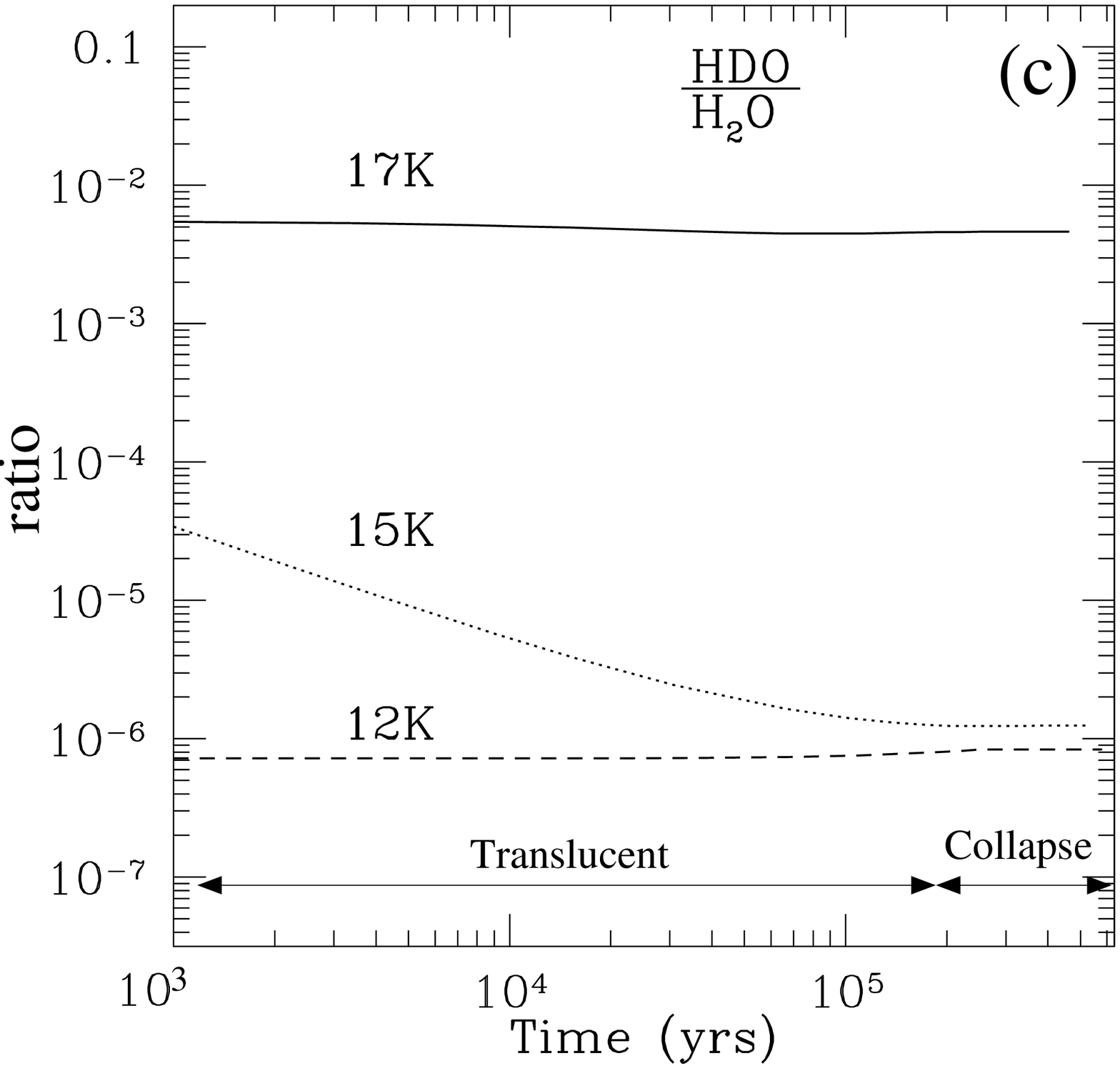}{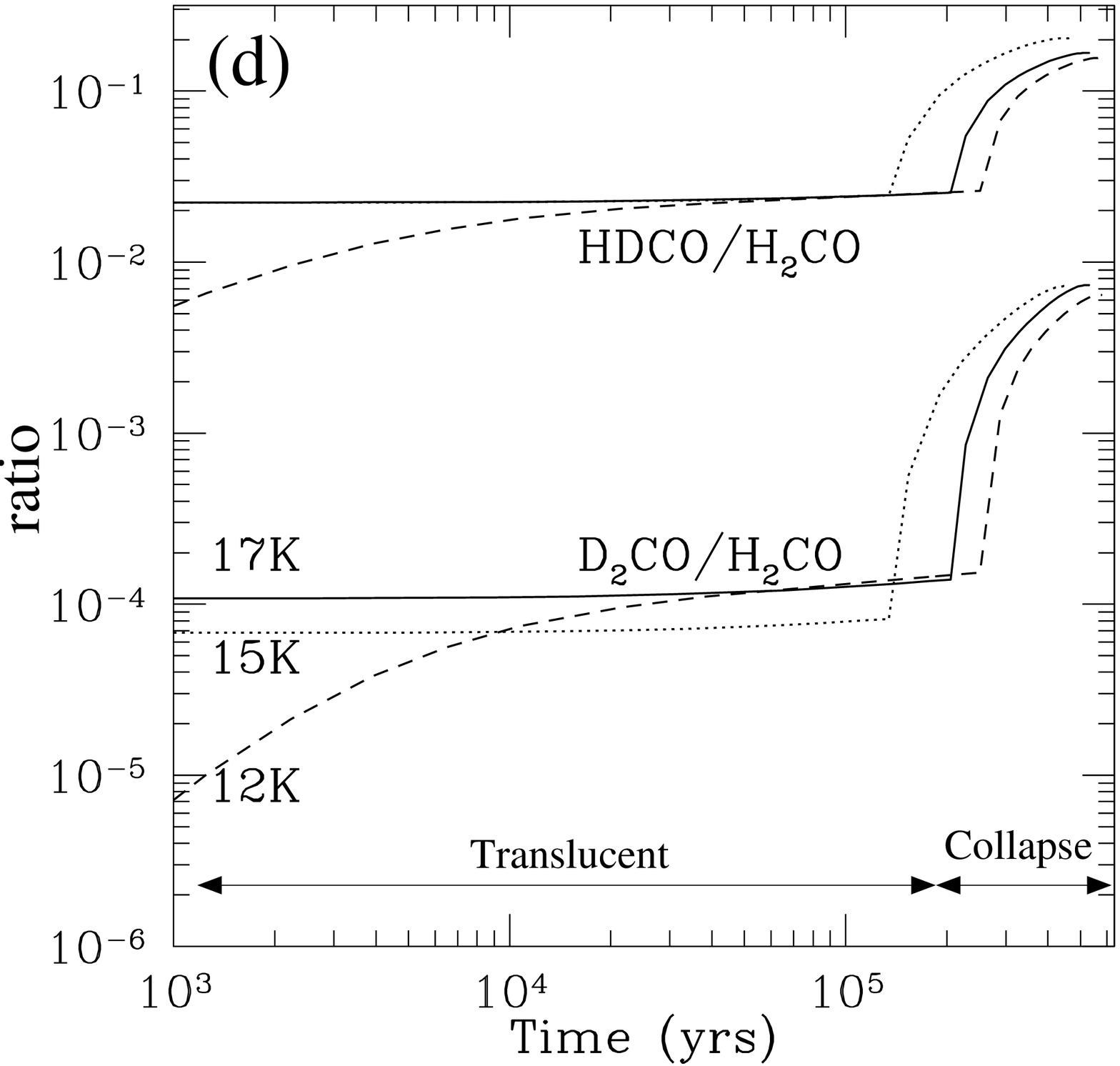} \\
\plottwo{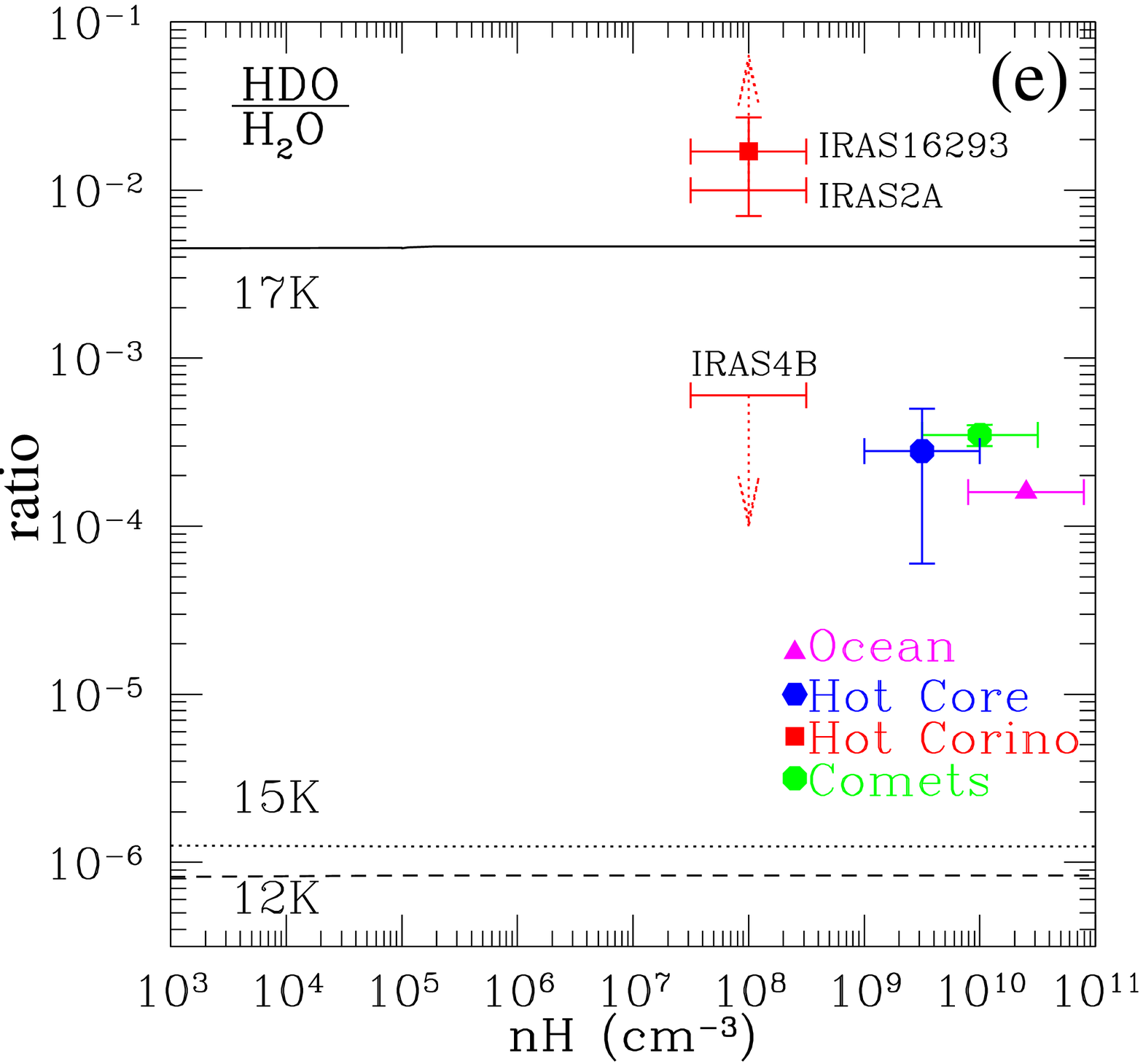}{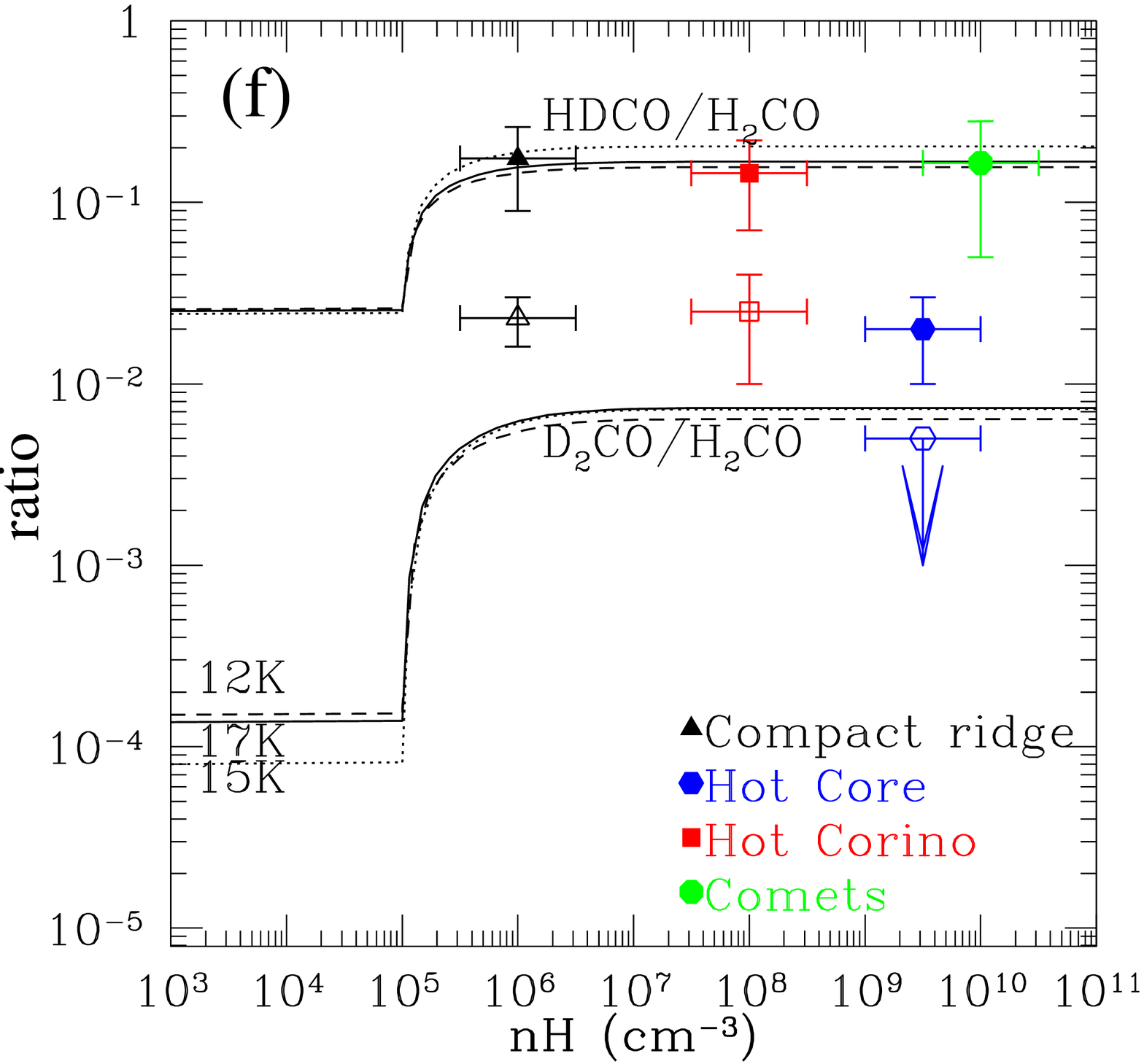}
\caption{H$_2$O, HDO, H$_2$CO, HDCO and D$_2$CO \rm{content on the dust during (1) the formation of ices in a translucent cloud and (2) the gravitational collapse of a cloud to form a star. Lines represent T$_{dust}$ =17~K (solid), 15~K (dotted) and 12~K (dashed)}. Top: surface coverage in monolayers ($\%$)  as function of time for H$_2$O and HDO (a) H$_2$CO, HDCO and D$_2$CO (b).  Middle:  deuterium fractionation of water (c) and formaldehyde (d). Bottom panels: model versus observations of HDO/H$_2$O (e) HDCO/H$_2$CO (filled symbols) D$_2$CO/H$_2$CO (open symbols) (f) towards several astrophysical objects as a function of total hydrogen density(Table 1).\rm}
\label{h2o}
\end{figure}

\end{document}